\documentclass[onecolumn,amsmath,amssymb,nofootinbib,aapt]{revtex4} 

\usepackage{amsmath}
\usepackage{graphicx}
\usepackage{graphics}
\usepackage{dcolumn}
\usepackage{bm}
\usepackage{hyperref}
\usepackage{xcolor}

\DeclareMathOperator\arctanh{arctanh}

\def\BEq{\begin{equation}}
\def\EEq{\end{equation}}
\def\BEqA{\begin{eqnarray}}
\def\EEqA{\end{eqnarray}}
\def\BEn{\begin{enumerate}}
\def\EEn{\end{enumerate}}
\def\BWT{\begin{widetext}}
\def\EWT{\end{widetext}}

\def\a{\alpha}

\begin{document}

\title{Modified Fronsdal coordinates for maximally extended Schwarzschild spacetime}

\author{Andrei Galiautdinov}
\email{ag1@uga.edu}
\affiliation{
Department of Physics and Astronomy, 
University of Georgia, Athens, Georgia 30602, USA
}

\date{\today}
\begin{abstract}
We introduce a coordinate system that complements the Kruskal--Szekeres extension. Like the standard construction, it covers the maximally extended Schwarzschild manifold in its entirety, while offering an additional advantage of expressing the areal radius as an explicit function of the new coordinates. Its main limitation, however, is that radial null geodesics are no longer represented as 45-degree lines in the Kruskal plane, making the causal structure more difficult to interpret. Nevertheless, the new system offers a compelling aesthetic trade-off: among all known maximally extended systems---including those of Kruskal--Szekeres, Israel, Fronsdal, Novikov, and Synge---it exhibits the highest degree of symmetry with respect to Schwarzschild's original $r$- and $t$-coordinate lines. It trades the regular pattern of Kruskal's light cones for a symmetric nesting arrangement of the two-dimensional spheres. The proposed extension sheds new light on the closely related Fronsdal's six-dimensional embedding construction, and clarifies the deep connection that exists between the most important implicit (Kruskal-Szekeres) and explicit (Israel's) procedures for maximal extension of the Schwarzschild geometry that is well known to those working in the field but rarely presented in textbooks on general relativity.
\end{abstract}

\maketitle


\section{Introduction}

The celebrated Kruskal-Szekeres procedure \cite{kruskal1960,szekeres1960} 
for maximal extension of the Schwarzschild vacuum,\footnote{Here we use 
the notation common in the educational literature on general relativity: 
$ds^2$ denotes the spacetime metric (line element, or interval), $(T,X)$ 
are the Kruskal--Szekeres coordinates, $r$ is the areal radius, $t$ is the 
Schwarzschild time measured by distant observers, $(\theta,\varphi)$ are 
the standard angular coordinates on the 2-sphere, and $r_{\rm s}=2M$ is 
the Schwarzschild radius, with $M$ the mass of the black hole expressed 
in geometrized units (meters). The areal radius $r$ is defined such that a 
2-sphere labeled by $r$ has surface area $4\pi r^2$; in general it does not 
coincide with the proper radial distance from any center---indeed, a center 
may not even exist in the spacetime (as in the case of a wormhole).}
\begin{align}
\label{eq:KSMetricStandard}
ds^2 
&=
\frac{4r_{\rm s}^3}{r} e^{-r/r_{\rm s}}
\left(
dT^2 -  dX^2
\right)
- r^2 d\Omega^2,
\quad
d\Omega^2 \equiv d \theta^2 + \sin^2 \theta \, d\varphi^2,
\end{align}
has its physical origins in Penrose's form \cite{penrose1965,penrose1969},
\begin{align}
\label{eq:EFPenrose}
v=t+\left(r+r_{\rm s}\ln\left| \frac{r}{r_{\rm s}}-1\right|\right),
\quad
u= t-\left(r+r_{\rm s}\ln\left| \frac{r}{r_{\rm s}}-1\right|\right),
\end{align}
of the Eddington-Finkelstein (EF) coordinates \cite{eddington1924, finkelstein1958}. 
The intimate connection between the two systems, $(T,X)$ and $(v,u)$, was 
clearly elaborated in Ref.\ \cite{MTW1973} by Misner, Thorne, and Wheeler, 
which undoubtedly helped both systems gain their tremendous popularity. 
The principal advantage of the Kruskal-Szekeres and other related doubly-null 
coordinates is that they always represent radial null rays as 45-degree lines in 
the Kruskal plane, making the causal structure of the underlying spacetime 
manifold immediately apparent.

For almost seventy years, the Kruskal--Szekeres extension has played a central 
role in general relativistic research. Its elegance, simplicity, and mathematical 
beauty have created a lasting legacy in the field, unlikely to be surpassed by 
any other extension system. Despite its enduring importance, however, the 
system has one notable drawback: it expresses the areal radius, $r$, through 
a pair of transcendental equations,
\begin{equation}
\label{eq:transcendentalKS}
e^{r/r_{\rm s}}\left(\frac{r}{r_{\rm s}} - 1\right) = X^2 - T^2,
\end{equation}
and 
\begin{equation}
\label{eq:transcendentalKSt}
\frac{t}{2r_{\rm s}} = 
\begin{cases} 
      \arctanh\left(\frac{T}{X}\right), & r>r_{\rm s}, \\
      \arctanh\left(\frac{X}{T}\right), & r<r_{\rm s},
   \end{cases}
\end{equation}
which prevents the metric coefficients from being written explicitly in terms 
of the adopted coordinates $(T, X)$. As a result, even routine computations, 
such as finding Christoffel symbols or curvature components, are beset with 
significant technical difficulties. This affects not only research but also how the 
subject is typically taught. While black holes and the Kruskal--Szekeres extension 
are routinely presented in general relativity courses, no other maximally extended 
system is usually introduced. Consequently, students who may be knowledgeable 
in the causal structure of black holes often find themselves unable to perform 
concrete, explicit analytical calculations, particularly those describing physics 
on the extended spacetime manifold. While we cannot reliably predict the 
direction in which observational research will take us tomorrow, a solid 
grounding in maximal extensions may one day become increasingly important, 
especially if more exotic objects, such as white holes, must eventually be taken 
into account. The following presentation is written with these considerations in 
mind and is intended, in part, to help bridge this gap in the training of students 
aiming to conduct creative research in general relativity.

The above, then, naturally raises the question: Can one construct an alternative, 
explicit coordinate system that likewise exhibits a compelling symmetry while 
also being suitable for practical calculations?

\begin{figure}
\includegraphics[angle=0,width=0.9\linewidth]{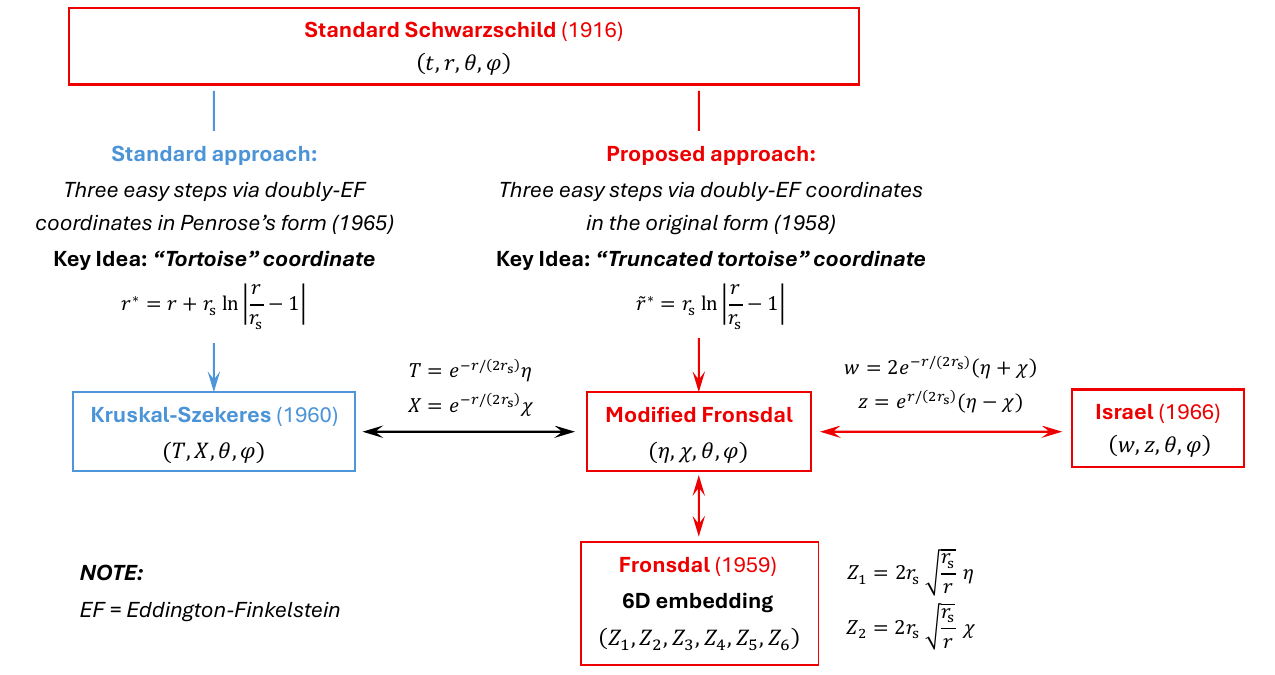}
\caption{ \label{fig:1}  
(Color online) Flowchart showing the relations among most important explicit 
(red) and implicit (blue) extension systems. The asymmetry of Israel's spacetime 
diagram (as depicted in Fig.\ 1 in Ref.\ \cite{israel1966}) is due to the opposing 
signs of the exponents appearing in the corresponding transformation equations. 
Generalization to other systems may be implemented via 
$r^{*}_{\rm general} = f(r) + r_{\rm s}\ln\left| \frac{r}{r_{\rm s}}-1\right|$ 
(however, see Sec. \ref{sec:simplifiedApproach} for a simpler method).
}
\end{figure}

It is evident from Eq.\ (\ref{eq:transcendentalKS}) that the main complication 
arises from the exponential prefactor. Therefore, the most straightforward 
approach would be to redefine the standard Kruskal-Szekeres coordinates 
so that this exponent is eliminated. This can be done quite easily, essentially 
by inspection. Readers who prefer this simplified approach may consult the 
flowchart in Fig.\ \ref{fig:1} (as well as Table \ref{tab:1}), and then proceed directly 
to our main equations, (\ref{eq:v04compact})-(\ref{eq:TandX}), and Section 
\ref{sec:simplifiedApproach}. 

For readers interested in a step-by-step construction starting from 
Schwarzschild’s original metric, we will offer a different strategy. We 
will retrace the usual sequence of transformations familiar from the 
standard Kruskal--Szekeres procedure, but with one important modification: 
instead of using the full radial tortoise coordinate,\footnote{The name 
comes from the singular transformation of the areal coordinate $r$ (see 
the chart in Fig.~\ref{fig:1}), which includes a logarithmically stretching 
contribution. This ingeniously chosen contribution allows the metaphorical 
tortoise to finally reach (and cross into!) the black hole (whereas in 
Schwarzschild coordinates the infalling photon geodesics appear stuck 
just outside of the horizon).} we will use its truncated version, as per 
Eddington and Finkelstein’s original proposal. This will allow us to carry 
out each step in a fully transparent and explicit manner, making the 
relationships between successive forms of the metric easy to understand. 
What we will essentially propose here is an  alternative path to the maximally 
extended Schwarzschild spacetime, which may be regarded as complementary 
to the traditional Kruskal-Szekeres formulation.

Before proceeding, let us interject a relevant historical remark.

One interesting explicit form of the maximally extended Schwarzschild metric 
was proposed by Israel in 1966 \cite{israel1966} 
(also, \cite{ehlers1973,pajerski1971,klosh1996}). Israel’s line element,
\begin{align}
\label{eq:israelMetric}
ds^2 
&=
r_{\rm s}^2
\left(
2dwdz - \frac{r_{\rm s}}{r} z^2dw^2
\right)
- r^2 d\Omega^2,
\quad
\frac{r}{r_{\rm s}} =1 -\frac{wz}{2},
\end{align}
was based solely on the ingoing Eddington-Finkelstein coordinates,\footnote{The 
connection is exhibited by $e^{v/(2r_{\rm s})}=w/2$, $r/r_{\rm s}=1-wz/2$, with 
$dv=2r_{\rm s}dw/w$, $dr=-(r_{\rm s}/2)(wdz+zdw)$, which upon substitution 
into the ingoing EF metric, $ds^2= \left(1-{r_{\rm s}}/{r}\right)dv^2-2dvdr- r^2 d\Omega^2$,  
produces Eq.\ (\ref{eq:israelMetric}).} which resulted in a highly asymmetric spacetime 
description. The pronounced lopsidedness of the corresponding spacetime diagram 
(as in Fig.\ 1 in Ref.\ \cite{israel1966}) may have contributed to the reasons why Israel’s 
important work did not receive the widespread attention it rightfully deserves. 
With this in mind, our approach can also be viewed as an attempt to symmetrize---and, 
in doing so, popularize---Israel’s extension by incorporating, in addition to the ingoing 
coordinates, the outgoing coordinates as well. What emerges from this ``symmetrizing'' 
construction is a four-dimensional line element that ties together all three historically 
important maximal extensions: the Kruskal-Szekeres implicit system, Israel’s explicit 
but asymmetric system, and the elegant, yet somewhat mysterious, six-dimensional 
pseudo-Euclidean isometric embedding
of the Schwarzschild spacetime,
\begin{align}
\label{eq:fronsdal6D}
ds^2 
&= 
dZ_1^2 - dZ_2^2 - dZ_3^2 - dZ_4^2 - dZ_5^2 -dZ_6^2,
\end{align}
introduced by Fronsdal in 1959 \cite{fronsdal1959}. It is therefore rather surprising 
that our metric (Eqs.\ (\ref{eq:v04}) and (\ref{eq:v04compact}) below) does not appear 
to have been previously written down or actually used. There were two early 
indications---in the well-known papers by Rindler \cite{rindler1966} and 
Synge \cite{synge1950})---suggesting what such a metric might look like. 
Apart from these, the only other work that touches upon the subject seems to 
be a recent pedagogical note by Unruh \cite{unruh2014}, in which a different 
approach to Fronsdal’s embedding is briefly discussed.

\begin{table}
\caption{\label{tab:1} Areal radius, $r$, in various coordinate systems 
considered in this article.}
\begin{ruledtabular}
\begin{tabular}{ccccc}
 {\bf Schwarzschild }& {\bf Kruskal-Szekeres}& {\bf Israel}& 
{\bf Fronsdal}&
{\bf Modified Fronsdal}
\\
$(t,r,\theta,\varphi)$ & $(T,X,\theta,\varphi)$ & $(w,z,\theta,\varphi)$  
& $(Z_1,Z_2,Z_3,Z_4,Z_5,Z_6)$ & $(\eta,\chi,\theta,\varphi)$
\\
\hline
\\
$r$ & $e^{r/r_{\rm s}}\left(\frac{r}{r_{\rm s}}-1\right)=X^2-T^2$ & 
$\frac{r}{r_{\rm s}} -1 = -\frac{wz}{2}$ &
$\frac{r}{r_{\rm s}}  =\left({1-\frac{Z_2^2-Z_1^2}{4r_{\rm s}^2}}\right)^{-1}$  &
$\frac{r}{r_{\rm s}} -1 = \chi^2-\eta^2$
\\
\\
\end{tabular}
\end{ruledtabular}
\end{table}

In the pages that follow, we will refer to the new metric as the modified 
Fronsdal extension. This name is appropriate because the final transformation 
(\ref{eq:fronsdalCoords1}) leading from our coordinates to those used by 
Fronsdal is remarkably simple, highlighting the direct algebraic relationship 
between the two. Our construction will preserve the essential geometric insight 
of Fronsdal’s original approach while reformulating it in a fully four-dimensional 
form. Unlike the original embedding, which requires six ambient dimensions 
and primarily serves as a geometric existence proof, our metric is immediately 
usable for calculations within the intrinsic spacetime. In this sense, our extension 
offers a streamlined and pedagogically transparent realization of essentially the 
same idea, thereby justifying the proposed terminology.

\section{The Metric}

Our detailed, physically motivated derivation of the modified Fronsdal extension 
for Schwarzschild geometry proceeds as follows.

Combining the original ingoing and outgoing Eddington-Finkelstein coordinates 
$(\tilde{v},\tilde{u})$ \cite{eddington1924, finkelstein1958},
\BEq
\label{eq:EFmodified}
\tilde{v}=t+r_{\rm s}\ln\left| \frac{r}{r_{\rm s}}-1\right|,
\quad
\tilde{u}=t-r_{\rm s}\ln\left| \frac{r}{r_{\rm s}}-1\right|,
\quad
t=(\tilde{v}+\tilde{u})/2,
\quad
r/r_{\rm s} = 
\begin{cases} 
      1 + e^{(\tilde{v}-\tilde{u})/(2r_{\rm s})}, & r>r_{\rm s}, \\
     1 - e^{(\tilde{v}-\tilde{u})/(2r_{\rm s})}, & r<r_{\rm s},
   \end{cases}
\EEq
with
\BEq
d\tilde{v}=dt+\frac{r_{\rm s}}{r}\frac{dr}{1- \frac{r_{\rm s}}{r}},
\quad
d\tilde{u}=dt-\frac{r_{\rm s}}{r}\frac{dr}{1- \frac{r_{\rm s}}{r}},
\EEq
we first convert the standard Schwarzschild line element, 
\begin{align}
\label{eq:v01introduction}
ds^2=\left(1-\frac{r_{\rm s}}{r}\right)dt^2
-\frac{dr^2}{1-\frac{r_{\rm s}}{r}}
- r^2 d\Omega^2,
\end{align}
to
\BEq
\label{eq:v02}
ds^2=\frac{1}{4}\left(1-\frac{r_{\rm s}}{r}\right)
\left[
(d\tilde{v}+d\tilde{u})^2-\left( \frac{r}{r_{\rm s}}  \right)^2 (d\tilde{v}-d\tilde{u})^2
\right]
- r^2 d\Omega^2.
\EEq
This step eliminates the most troubling metric coefficient $g_{rr}$, but places 
$r=r_{\rm s}$ at $\tilde{v}-\tilde{u}=-\infty$.\footnote{Which corresponds to 
setting $e^{(\tilde{v}-\tilde{u})/(2r_{\rm s})}=0$ in Eq.\ (\ref{eq:EFmodified}).}  
To bring the horizon back from coordinate 
infinity, we employ the re-scaling trick \cite{martel2001, poisson2009},
\BEq
\tilde{V}=e^{\tilde{v}/(2r_{\rm s})}, 
\quad 
\tilde{U}=\mp e^{-\tilde{u}/(2r_{\rm s})},
\quad
d\tilde{v}=2r_{\rm s}{d\tilde{V}}/{V},
\quad
d\tilde{u}=-2r_{\rm s}{d\tilde{U}}/{\tilde{U}},
\EEq
with $\mp$ referring to $r>r_{\rm s}$ and $r<r_{\rm s}$, 
respectively, which gives
\BEq
\tilde{V}\tilde{U}=-\frac{r}{r_{\rm s}}\left(1-\frac{r_{\rm s}}{r}\right), \quad r>0,
\EEq
valid for any $r$. This allows elimination of the prefactor 
$\left(1-{r_{\rm s}}/{r}\right)$ in (\ref{eq:v02}), leading to
\begin{align}
\label{eq:v03}
ds^2 
&=
r_{\rm s}^2
\left[
2\left(
\frac{r}{r_{\rm s}}+\frac{r_{\rm s}}{r}
\right) d\tilde{V}d\tilde{U}
-\left(1+ \frac{r_{\rm s}}{r} \right)
\left(\tilde{U}^2d\tilde{V}^2+\tilde{V}^2d\tilde{U}^2\right)
\right]
- r^2 d\Omega^2,
\end{align}
where
\BEq
\label{eq:trVU}
{r}/{r_{\rm s}}=1-\tilde{V}\tilde{U},
\quad
e^{t/r_{\rm s}}=\mp {\tilde{V}}/{\tilde{U}}.
\EEq
The resulting metric may be viewed as the symmetrized version of 
Israel's line element introduced in Ref.\ \cite{israel1966} (also see some 
interesting discussions in Refs.\ 
\cite{lake2006,lake2010,roken2022,bisson2023,cederbaum2024}.) 
This can be seen by exhibiting the transformation equations linking 
the two systems,
\BEq
\label{eq:newToIsrael}
w=2e^{r/(2r_{\rm s})}\tilde{V},
\quad
z=e^{-r/(2r_{\rm s})}\tilde{U},
\EEq
which, after some algebra, recovers Eq.\ (\ref{eq:israelMetric}). Notice that 
it is the presence of the opposite signs in the exponents in (\ref{eq:newToIsrael}) 
which makes Israel's spacetime diagram distinctly asymmetric 
(compare to (\ref{eq:TandX}) in the Kruskal case below).
Finally, an additional coordinate transformation,
\BEq
\label{eq:ct02}
\tilde{V}=\eta+\chi, 
\quad 
\tilde{U}=\eta-\chi,
\EEq
brings the metric to the form,
\begin{align}
\label{eq:v04}
ds^2 
&=
\frac{4r_{\rm s}^3}{r}
\left\{
\left[ 1-\eta^2\left(1+\frac{r}{r_{\rm s}}\right)\right] d\eta^2
+ 2 \eta \chi \left(1+\frac{r}{r_{\rm s}}\right) d\eta d\chi
-\left[ 1+\chi^2\left(1+\frac{r}{r_{\rm s}}\right)\right] d\chi^2
\right\}
- r^2 d\Omega^2,
\end{align}
or, somewhat more compactly,
\begin{align}
\label{eq:v04compact}
ds^2 
&=
\frac{4r_{\rm s}^3}{r}
\left[
d\eta^2 -  d\chi^2
- \left(1+\frac{r}{r_{\rm s}}\right) 
\left(\eta d\eta - \chi d\chi \right)^2
\right]
- r^2 d\Omega^2,
\end{align}
where
\BEq
\label{eq:trTX}
\frac{r}{r_{\rm s}}=1+\chi^2-\eta^2,
\quad
\frac{t}{2r_{\rm s}} = 
\begin{cases} 
      \arctanh\left(\frac{\eta}{\chi}\right), & r>r_{\rm s}, \\
      \arctanh\left(\frac{\chi}{\eta}\right), & r<r_{\rm s},
   \end{cases}
\EEq
which should be compared to Eqs.\ (\ref{eq:transcendentalKS}) 
and (\ref{eq:transcendentalKSt}), and
\begin{align}
\label{eq:TXtr}
r>r_{\rm s}:
\begin{cases} 
\eta = \left|\frac{r}{r_{\rm s}}-1\right|^{1/2}
\sinh\left(\frac{t}{2r_{\rm s}}\right),
\\
\chi = \left|\frac{r}{r_{\rm s}}-1\right|^{1/2}
\cosh\left(\frac{t}{2r_{\rm s}}\right),
\end{cases}
\quad
r<r_{\rm s}:
\begin{cases} 
\eta = \left|\frac{r}{r_{\rm s}}-1\right|^{1/2}
\cosh\left(\frac{t}{2r_{\rm s}}\right),
\\
\chi = \left|\frac{r}{r_{\rm s}}-1\right|^{1/2}
\sinh\left(\frac{t}{2r_{\rm s}}\right),
\end{cases}
\end{align}
where both branches of the square root must be taken into account.
\begin{figure}
\includegraphics[angle=0,width=1.00\linewidth]{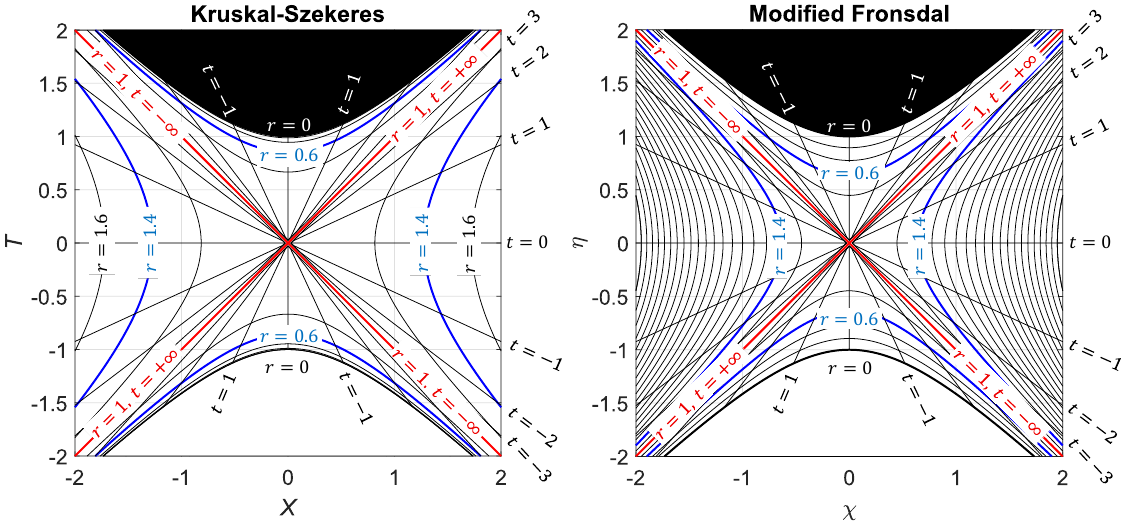}
\caption{ \label{fig:2}  
(Color online) Comparison of the Kruskal and modified Fronsdal diagrams. 
Each point in either diagram represents a two-dimensional sphere of areal 
radius $r$. The $r=\text{constant}$ lines are plotted in steps of $0.2\,r_{\rm s}$ 
and form hyperbolas in both cases. The factor $r_{\rm s}$ on the labels of curves 
is omitted for clarity. To emphasize the enhanced symmetry of the modified 
Fronsdal diagram, a representative pair of curves is highlighted in blue color 
($r = 0.6\,r_{\rm s}$ and $1.4\,r_{\rm s}$), while the remaining $r=\text{constant}$ 
curves are shown in black. Straight lines of constant $t$ passing through the 
origin are plotted for $t/r_{\rm s} = -\infty, -3, -2, -1, 0, 1, 2, 3, \infty$, with the 
red lines representing the white and black hole horizons ($r = r_{\rm s}$, 
$t = \pm \infty$). The usual Kruskal--Szekeres coordinates $(T, X)$ are related 
to the modified Fronsdal coordinates $(\eta, \chi)$ via Eq.\ (\ref{eq:TandX}), 
preserving the ratio $T/X = \eta/\chi$.
}
\end{figure}

This indicates a simple connection between Kruskal's original and the modified 
coordinates,
\BEq
\label{eq:TandX}
T =e^{r/(2r_{\rm s})}\eta, 
\quad 
X=e^{r/(2r_{\rm s})}\chi,
\EEq 
which will be exploited in what follows. The six-dimensional isometric 
embedding is then immediately recovered by introducing the Fronsdal 
coordinates via
\begin{align}
\label{eq:fronsdalCoords1}
Z_1 = 2r_{\rm s}\sqrt{\frac{r_{\rm s}}{r}} \, \eta,
\quad
Z_2 = 2r_{\rm s}\sqrt{\frac{r_{\rm s}}{r}} \, \chi,
\end{align}
and
\begin{align}
\label{eq:fronsdalCoords2}
Z_3 = \int^r \sqrt{
\frac{r_{\rm s}}{r} + \frac{r_{\rm s}^2}{r^2} + \frac{r_{\rm s}^3}{r^3}} \, dr,
\quad
Z_4 &= r \sin \theta \cos \varphi,
\quad
Z_5 = r \sin \theta \sin \varphi,
\quad
Z_6 = r \cos \theta,
\end{align}
with
\BEq
r
=
\frac{r_{\rm s}}{1-\frac{Z_2^2-Z_1^2}{4r_{\rm s}^2}}
=
\sqrt{Z_4^2+Z_5^2+Z_6^2},
\EEq
in terms of which the line element assumes the famed pseudo-Euclidean 
form (\ref{eq:fronsdal6D}). One may appreciate how naturally the transformation 
equations (\ref{eq:fronsdalCoords1}) and (\ref{eq:fronsdalCoords2}) defining 
Fronsdal's embedding have emerged in our derivation (cf.\ Ref.\ \cite{fronsdal1959}; 
also \cite{sassi1988}).

The spacetime diagram corresponding to (\ref{eq:trTX}) is depicted in 
Fig.\ \ref{fig:1} (right panel). Notice that in our  extension the constant-$t$ lines 
of the original Schwarzschild geometry, unlike in Israel's, are straight, passing 
through the origin, and are similar to Kruskal's. The constant-$r$ lines are also 
beautifully distributed, forming a highly symmetric nesting arrangement surrounding 
the horizon (cf.\ Gullstrand-Painlev\'{e}-type extensions recently discussed in 
Ref.\ \cite{lemos2021}; also  \cite{martel2001} and \cite{unruh2014}).

\section{The Causal Structure}

\subsection{Radial Null geodesics}
\label{sec:radialNullGeodesics}

\begin{figure}
\includegraphics[angle=0,width=0.5\linewidth]{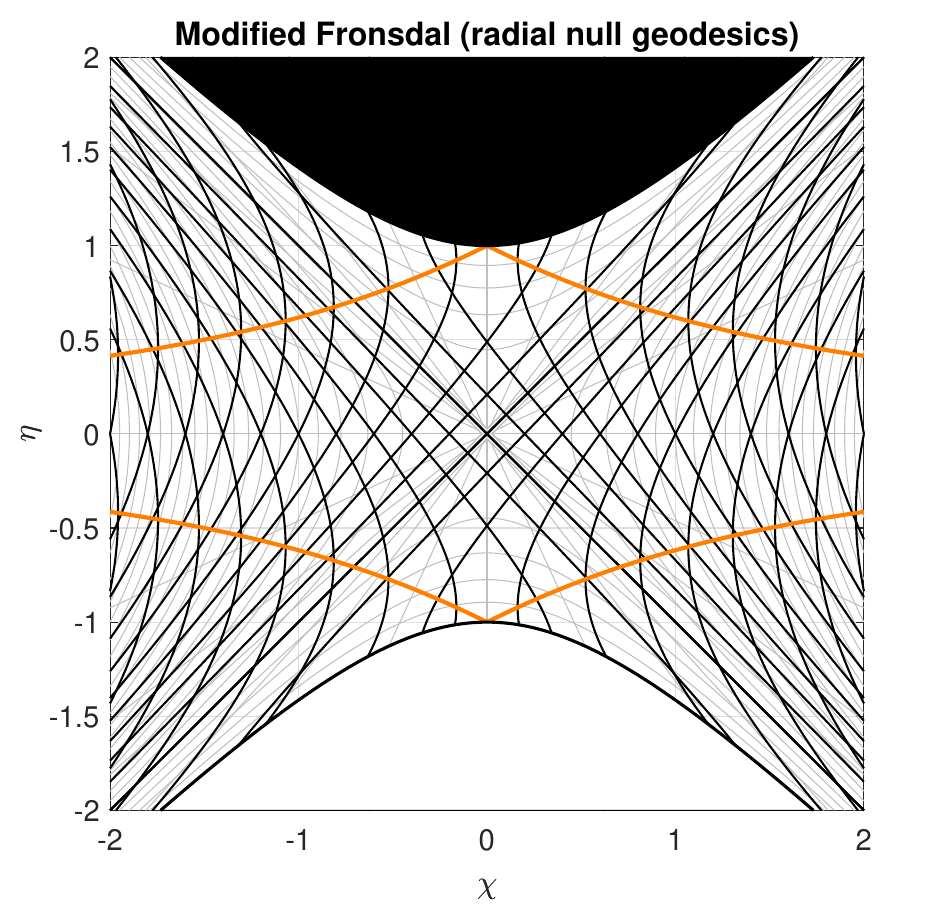}
\caption{ \label{fig:3}  
(Color online) Radial null geodesics (black curves) in modified 
Fronsdal coordinates plotted on the basis of 
Eq.\ (\ref{eq:nullGeodesics}), where $\chi_0$ varies in steps of $0.2$. 
The meaning of the curves plotted in orange is elaborated upon 
in Sec.\ \ref{sec:hypersurfaces}. Notice characteristic 
Finkelsteinian collapse of the light cones as they approach the 
singularity. Compare to the light-cone structure depicted 
in Fig.\ 2 of Fronsdal's original paper \cite{fronsdal1959}.
}
\end{figure}

In the modified Fronsdal coordinates, $(\eta,\chi)$, the ingoing and outgoing 
radial null geodesics (plotted in Fig.\ \ref{fig:3}) are described by the equation,
\BEq
\label{eq:nullGeodesics}
\eta \pm \chi = Ce^{(\eta^2-\chi^2)/2},
\quad
C=\chi_0 e^{\chi_0^2/2},
\EEq
where $\chi_0$ is the value of the $\chi$ coordinate at $\eta=0$. This may be 
seen most easily by differentiating (\ref{eq:TandX}),
\BEq
{dT}/{dX}=\mp 1
=
\left({\eta}+2r_{\rm s}\,{d\eta}/{dr}\right)/\left({\chi}+2r_{\rm s}\,{d\chi}/{dr}\right),
\EEq 
whose solution is
\BEq
\eta \pm \chi = \tilde{C}e^{-r/(2r_{\rm s})},
\EEq
which is equivalent to (\ref{eq:nullGeodesics}) on the basis of (\ref{eq:trTX}).

\subsection{Gullstrand-Painlev\'{e} ``raindrops''}

\begin{figure}
\includegraphics[angle=0,width=0.5\linewidth]{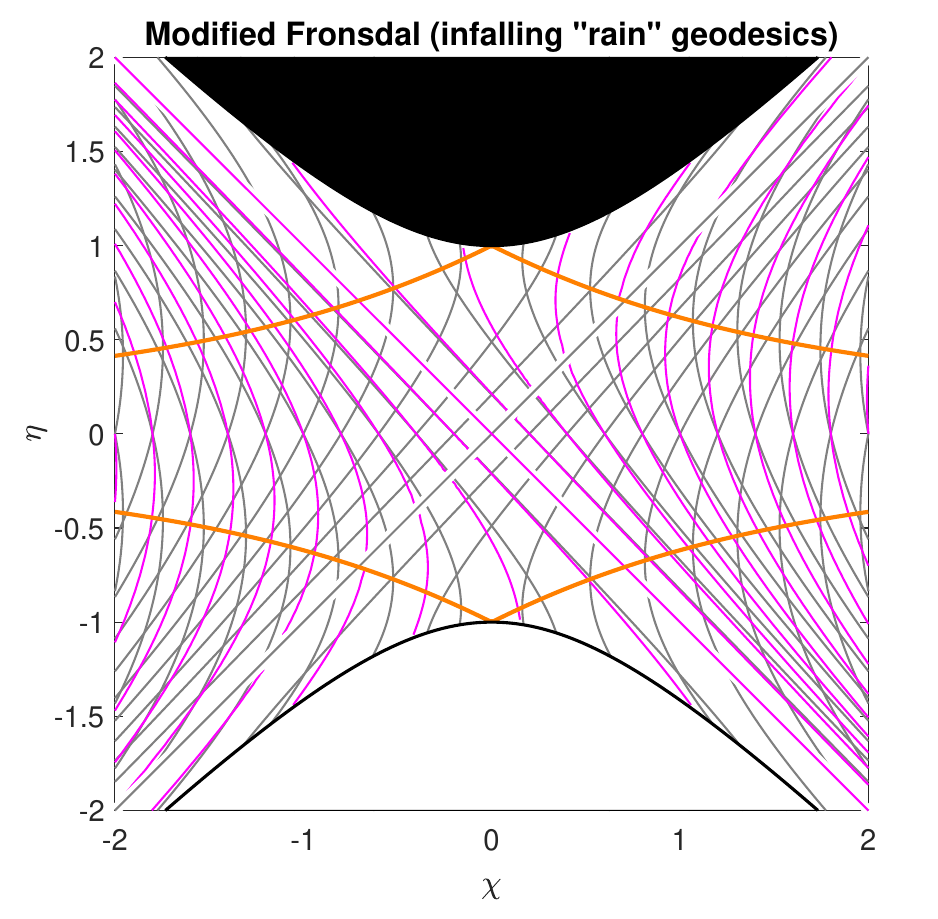}
\caption{ \label{fig:4}  
(Color online) Infalling Gullstrand-Painlev\'{e} geodesics (magenta curves) 
superimposed on radial null geodesics (light gray curves) 
plotted on the basis of Eq.\ (\ref{eq:timelikeGeodesic2}), 
where $\chi_0$ varies in steps of $0.2$. 
}
\end{figure}

By Gullstrand-Painlev\'{e} raindrops we mean a collection 
of point-like massive probes radially infalling from rest at 
spatial infinity. In modified Fronsdal geometry, the corresponding 
infalling timelike geodesics are described by the equation,
\BEq
\label{eq:timelikeGeodesic1}
\frac{d\eta}{d\chi}
=
\frac{\chi \left({r}/{r_{\rm s}}\right)^{3/2}-\eta}
{\eta \left({r}/{r_{\rm s}}\right)^{3/2}-\chi},
\EEq
or,
\BEq
\label{eq:timelikeGeodesic2}
\frac{\eta+\chi}{\eta-\chi}
=
\mp C \left|
\frac{\left({r}/{r_{\rm s}}\right)^{1/2}+1}
{\left({r}/{r_{\rm s}}\right)^{1/2}-1}
\right|
\exp\left\{
-\left[
\frac{2}{3}\left(\frac{r}{r_{\rm s}}\right)^{3/2}
+2\left(\frac{r}{r_{\rm s}}\right)^{1/2}
\right]
\right\},
\quad
\frac{r}{r_{\rm s}} = 1+\chi^2-\eta^2,
\EEq
where
\BEq
C 
= 
{
\left|
\frac{\left({r_0}/{r_{\rm s}}\right)^{1/2}-1}
{\left({r_0}/{r_{\rm s}}\right)^{1/2}+1}
\right|
}
{
\exp\left\{\frac{2}{3}\left(\frac{r_0}{r_{\rm s}}\right)^{3/2}
+2\left(\frac{r_0}{r_{\rm s}}\right)^{1/2}\right\}
},
\quad
\frac{r_0}{r_{\rm s}}=1+\chi_0^2,
\EEq
with $\mp$ referring to $r>r_{\rm s}$ and $r<r_{\rm s}$, 
respectively. In the above, $\chi_0$ is the value of the 
$\chi$-coordinate at $\eta=0$ (see Fig.\ \ref{fig:4}). 
The proof (left as a straightforward exercise) of Eqs.\ (\ref{eq:timelikeGeodesic1}) and 
(\ref{eq:timelikeGeodesic2}) boils down to rewriting various well known
results pertaining to radially infalling geodesics of the standard 
Schwarzschild metric in terms of the newly introduced coordinates 
$(\eta,\chi)$. The extended solution then, in addition, automatically 
describes the motion of the (formally regarded as infalling) 
``raindrops'' that come to rest at the spatial infinity of the 
``other'' universe after being emitted by the past singularity (white hole).

\subsection{The nature of the constant-$\eta$ and constant-$\chi$ hypersurfaces}

\label{sec:hypersurfaces}

Referring back to Fig.\ (\ref{fig:3}), let us take a closer look at the central, 
diamond-shaped region bounded by the curves plotted in orange. These 
particular curves are singled out because they are defined by the condition
\begin{equation}
\label{eq:diamondBoundaryModFronsdal}
\eta = \pm \frac{1}{\sqrt{1+\frac{r}{r_{\rm s}}}},
\end{equation}
or, equivalently,
\begin{equation}
\eta = \pm \sqrt{1+\frac{\chi^2}{2}-\sqrt{\chi^2 \left(1+\frac{\chi^2}{4}\right)}},
\end{equation}
which correspond precisely to the loci where $g_{00}=0$. In other words, 
they mark the transition between timelike and spacelike behavior of the 
$\eta$-lines (of constant $\chi$) and pass through the ``turning points'' of 
various null geodesics. Within this region,
\begin{equation}
g_{00} = \frac{4r_{\rm s}^3}{r}
\left[ 1-\eta^2\left(1+\frac{r}{r_{\rm s}}\right)\right] d\eta^2 > 0 
\quad \text{(positive)}, 
\end{equation}
which implies that the bounded segments of $\eta$-lines  threading the 
diamond are timelike and may be interpreted as the worldlines of certain 
temporarily existing, physically realizable, accelerated (except for the central one) 
local observers. Outside this region, the $\eta$-lines are spacelike (tachyonic) 
and therefore cannot be associated with physical observers.

\begin{figure}
\includegraphics[angle=0,width=1\linewidth]{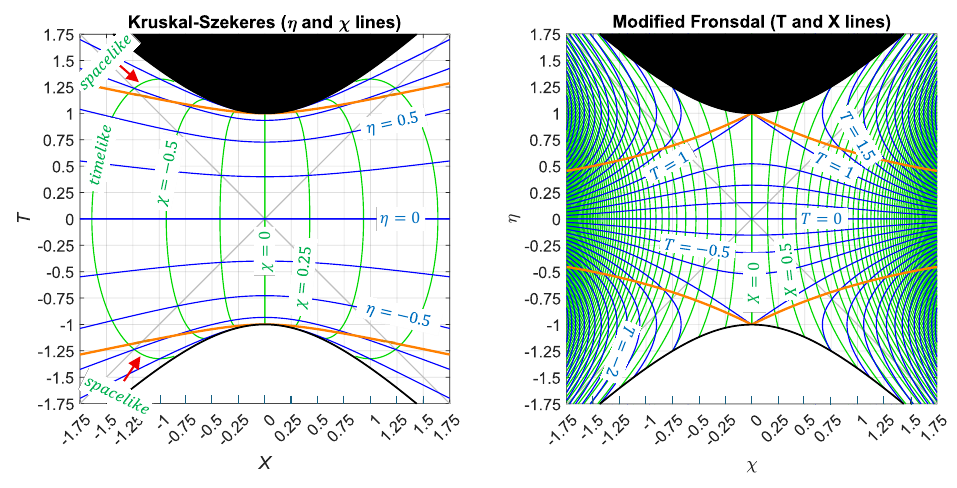}
\caption{ \label{fig:5}  
(Color online) Left panel: the constant-$\eta$ (blue) and constant-$\chi$ (green) 
lines in the Kruskal-Szekeres coordinates $(T,X)$. Right panel: Kruskal's constant-$T$ 
(blue) and constant-$X$ (green) lines in the modified Fronsdal coordinates $(\eta,\chi)$. 
Compare to Figs.\ 1 and 2 in Ref.\ \cite{martel2001} and Fig.\ 2 in Ref. \cite{unruh2014}.
}
\end{figure}

These considerations prompt us to re-draw the $\eta$- and $\chi$-lines, as 
well as the diamond itself, in the original Kruskal plane. Note that the diamond’s 
boundary is now described by
\begin{equation}
1 + e^{-|X/T|}\left(X^2 - T^2\right) = \pm \frac{X}{T},
\end{equation}
which follows from Eqs.\ (\ref{eq:diamondBoundaryModFronsdal}), (\ref{eq:TandX}), 
and (\ref{eq:transcendentalKS}). The results are depicted in Fig.\ \ref{fig:5} (left panel), 
where, for completeness, we have also included the reciprocal plot showing the 
$T$- and $X$-lines in $(\eta, \chi)$ coordinates (right panel).

Notice again that each of the $\eta$-lines (lines of constant $\chi$) consists of a 
central timelike segment contained within the diamond, flanked by spacelike 
segments on either side. These outer portions, when represented in the original 
Kruskal plane, appear to extend backward in the Kruskal time coordinate $T$, 
making their spacelike character particularly evident (cf.\ the cases of 
Gullstrand-Painlev\'{e} in Ref.\ \cite{martel2001} and Synge in Ref. \cite{unruh2014}).

In addition, the $\chi$-lines (lines of constant $\eta$) provide a natural slicing 
of the spacetime into spacelike hypersurfaces. This foliation will serve as a useful 
tool in the next section, where we examine the dynamical interpretation of the 
Schwarzschild wormhole.

\section{Application to Wormhole Dynamics}
\label{sec:embedding}

Let us now apply Eq.\ (\ref{eq:v04}) to the Einstein--Rosen bridge \cite{einstein1935}, 
also known as the Schwarzschild wormhole. The classic analysis by Fuller and 
Wheeler \cite{fuller1962} employs Kruskal’s implicit coordinate system, which 
makes the description rather elaborate. While their treatment has become a 
standard reference in the literature and reveals many essential features of the 
wormhole's geometry, the associated spacetime foliation---based on the auxiliary 
parameter $\gamma = T/(4 + X^2)^{1/2}$---was, by their own admission, 
``quite arbitrarily selected.'' Although mathematically consistent, this slicing 
does not arise naturally from the coordinates themselves and appears somewhat 
artificial. In contrast, our approach uses an explicit coordinate system, in which 
the foliation by constant-$\eta$ hypersurfaces (the $\chi$-lines) emerges directly 
from the geometry and provides a more transparent, coordinate-adapted setup 
for analyzing wormhole dynamics. In this sense, the proposed approach 
complements the classic treatment by offering an alternative, geometrically 
motivated framework for exploring wormhole dynamics.

\begin{figure}
\includegraphics[angle=0,width=0.6\linewidth]{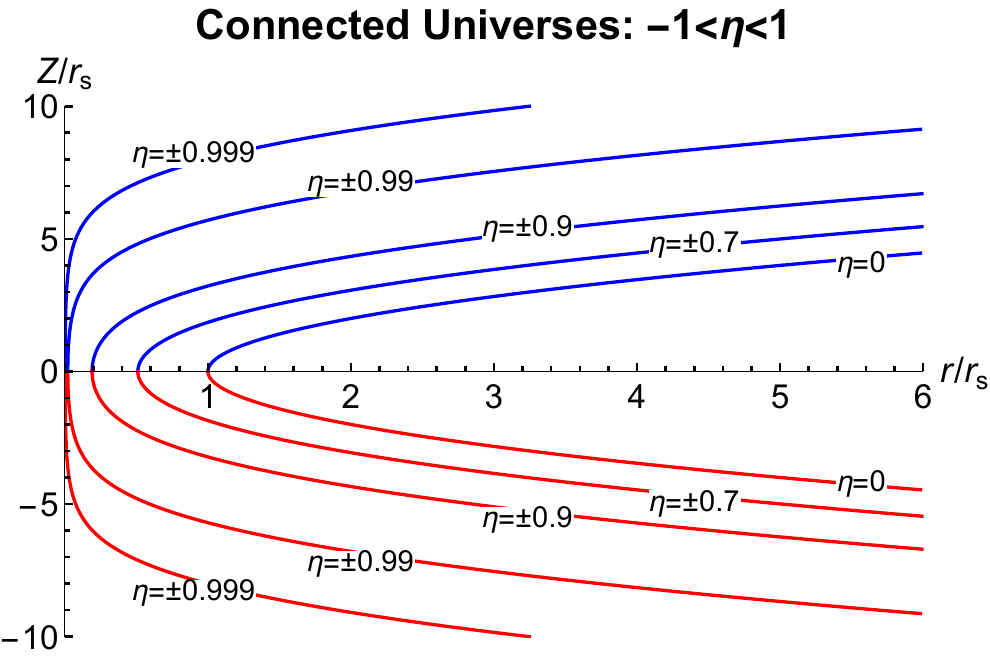}
\caption{ \label{fig:6}  
(Color online) The $\eta$-dependent Flamm's embedding of Schwarzchild's 
wormhole as described by Eq.\ (\ref{eq:embeddingExplicitFormula}) 
for $0\leq |\eta| < 1$. ``Our'' universe is depicted in blue (with $Z/r_{\rm s} > 0)$, the 
``other'' universe is depicted in red (with $Z/r_{\rm s} < 0)$. The throat is maximally 
open at $\eta=0$ 
($r_{\rm throat} = r_{\rm s}$). 
}
\end{figure}

\begin{figure}
\includegraphics[angle=0,width=0.6\linewidth]{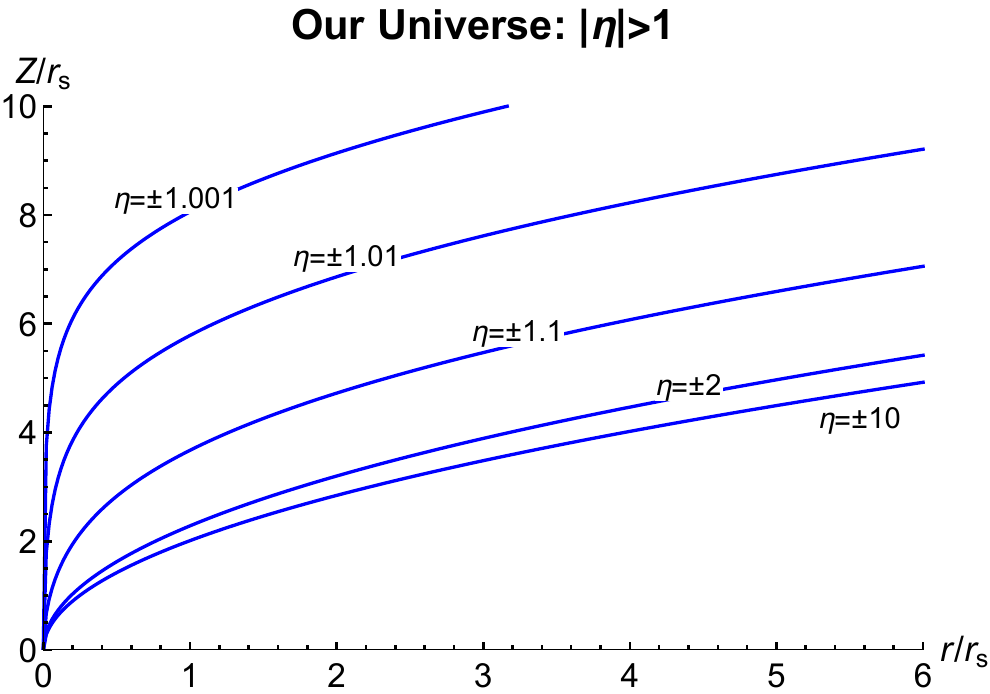}
\vskip25pt
\includegraphics[angle=0,width=0.6\linewidth]{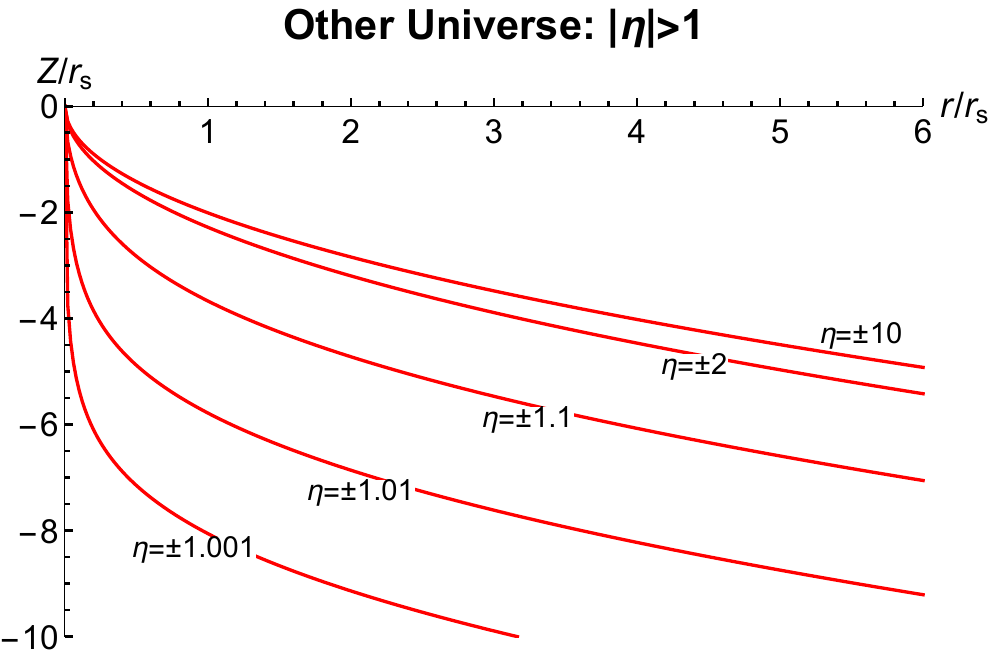}
\caption{\label{fig:7}  
(Color online) Evolution of the two disconnected universes at $|\eta|>1$,
as described by Eq.\ (\ref{eq:embeddingExplicitFormula}). 
Notice that the cusps of the corresponding embeddings 
settle on certain limiting shapes as $\eta \rightarrow \pm\infty$, 
which in this figure are virtually indistinguishable from those 
at $\eta=\pm 10$. The figure may give the misleading impression that the 
two disconnected universes touch at $Z = 0$, which is not the case. For $|\eta| > 1$, 
the universes are in fact spatially separated in the $\chi$ (or $X$) direction.
}
\end{figure}

Referring to Eq.\ (\ref{eq:v04}), we see that at any fixed $\eta$, the wormhole's 
spatial geometry is described by the line element
\begin{equation}
d\ell^2
=
\frac{4r_{\rm s}^3}{r}
\left[1+\chi^2\!\left(1+\frac{r}{r_{\rm s}}\right)\right] d\chi^2
+ r^2 d\Omega^2,
\end{equation}
with
\[
\chi^2 = \frac{r}{r_{\rm s}} - 1 + \eta^2, 
\qquad
2\chi\, d\chi = \frac{dr}{r_{\rm s}}, 
\qquad
d\chi^2 = \frac{dr^2}{4r_{\rm s}^2 \chi^2}.
\]
Its Flamm embedding in Euclidean space \cite{flamm1916},
\begin{equation}
d\ell^2 = dr^2 + dZ^2 + r^2 d\Omega^2,
\end{equation}
is therefore given by
\begin{equation}
\label{eq:embeddingMetric}
dZ^2 = \frac{1+\eta^2/\rho}{\rho - 1 + \eta^2}\, dr^2,
\qquad
\rho \equiv \frac{r}{r_{\rm s}},
\end{equation}
or, equivalently,\footnote{Here, $\rho_{0}$ is chosen so that the minimal areal 
radius $\rho_{0} \equiv r_{\rm min}/r_{\rm s}$ on each of the Flamm plots 
(Fig.\ \ref{fig:6} and Fig.\ \ref{fig:7}) corresponds to $Z=0$. For $|\eta| < 1$ 
(connected universes), the condition $dZ/dr = \pm \infty$, which defines 
$r_{\rm min}$, then implies from Eq.\ (\ref{eq:embeddingMetric}) that 
$\rho_{0} - 1 + \eta^2 = 0$, i.e., $\rho_{0} = 1 - \eta^2$. For $|\eta| > 1$ 
(disconnected universes), we set $\rho_{0}=0$ to ensure that the integrand 
never becomes imaginary.}
\begin{equation}
\label{eq:embeddingExplicitFormula}
\frac{Z}{r_{\rm s}}
=
\pm \int\limits_{\rho_{0}}^{\rho}
\sqrt{\frac{1+\chi^2/\xi}{\xi - 1 + \eta^2}}\, d\xi,
\qquad
\rho_{0}
= 
\begin{cases} 
 		1-\eta^2, & 0 \leq |\eta| < 1, \\[6pt]
               0, & |\eta| > 1, 
   \end{cases}
\end{equation}
whose $\eta$-dependent profiles are plotted in Fig.\ \ref{fig:6} 
and Fig.\ \ref{fig:7} (cf., e.g., Ref.\ \cite{chyba1982} and the more 
recent Ref.\ \cite{collas2012}).

Notice that for $\eta$ such that $0 \leq |\eta| < 1$, 
Eq.\ (\ref{eq:embeddingExplicitFormula}) may be written in the alternative form
\begin{align}
\label{eq:altFlamm}
0 \leq |\eta| < 1:\quad
\frac{Z}{r_{\rm s}}
&=
\pm
2\left[
\sqrt{(\rho-1+\eta^2)\!\left(1+\frac{\eta^2}{\rho}\right)}
-
\eta \int \limits_{\eta/\sqrt{1-\eta^2}}^{\eta/\sqrt{\rho}}
\sqrt{\frac{1-\!\left(\frac{1}{\eta^2}-1\right)\zeta^2}{1+\zeta^2}} \, d\zeta
\right].
\end{align}
If needed, the integral in square brackets may be expressed 
in terms of the hyperbolic version of the incomplete elliptic integral 
of the second kind,
\begin{equation}
\tilde{E}(\phi,k)
\equiv
\int\limits_{0}^{\phi}\sqrt{1-k^2\sinh^2\psi}\, d\psi
=
\int\limits_{0}^{\sinh\phi}
\sqrt{\frac{1-k^2\zeta^2}{1+\zeta^2}}\, d\zeta .
\end{equation}
Either way, at $\eta=0$ we recover the standard result for the maximally
open throat,
\begin{equation}
\eta=0:\quad 
\frac{Z}{r_{\rm s}} = 2\,\sqrt{\frac{r}{r_{\rm s}} - 1}.
\end{equation}

The proof of Eq.\ (\ref{eq:altFlamm}) proceeds as follows.

Our strategy is to recast the integrand in (\ref{eq:embeddingExplicitFormula}) 
into a form that naturally separates into two contributions: one that is the 
exact differential of an algebraic term, and another that integrates to the 
elliptic-integral--like expression.
First, observe that
\begin{align}
\sqrt{\frac{1+\eta^2/\xi}{\xi - 1+\eta^2}} 
&=
\left[
\sqrt{\frac{1+\eta^2/\xi}{\xi - 1+\eta^2}}
-
\sqrt{\frac{\xi - 1+\eta^2}{\xi + \eta^2}}\,\frac{\eta^2}{\xi^{3/2}}
\right]
+
\sqrt{\frac{\xi - 1+\eta^2}{\xi + \eta^2}}\,\frac{\eta^2}{\xi^{3/2}}
\nonumber \\
&=
\left[
\frac{1+\eta^2/\xi}
{\sqrt{(\xi - 1+\eta^2)\!\left(1+\eta^2/\xi\right)}}
+
\frac{(\xi - 1+\eta^2)(-\eta^2/\xi^2)}
{\sqrt{(\xi - 1+\eta^2)\!\left(1+\eta^2/\xi\right)}}
\right]
\nonumber \\
\label{eq:proofE1}
&\quad
-
\sqrt{\frac{1-\!\left(\tfrac{1}{\eta^2}-1\right)(-\eta/\xi^{1/2})^2}
{1+(-\eta/\xi^{1/2})^2}}\,\left(-\frac{\eta^2}{\xi^{3/2}}\right).
\end{align}
Next, consider differentials,
\begin{align}
\label{eq:proofE2}
d \sqrt{(\xi - 1+\eta^2)\!\left(1+\frac{\eta^2}{\xi}\right)}
&=
\frac{1}{2}\,
\frac{d\!\left[(\xi - 1+\eta^2)\!\left(1+\tfrac{\eta^2}{\xi}\right)\right]}
{\sqrt{(\xi - 1+\eta^2)\!\left(1+\tfrac{\eta^2}{\xi}\right)}}
\nonumber \\
&=
\frac{1}{2}\left[
\frac{1+\eta^2/\xi}
{\sqrt{(\xi - 1+\eta^2)\!\left(1+\eta^2/\xi\right)}}
+
\frac{(\xi - 1+\eta^2)(-\eta^2/\xi^2)}
{\sqrt{(\xi - 1+\eta^2)\!\left(1+\eta^2/\xi\right)}}
\right] d\xi ,
\end{align}
and
\begin{align}
\label{eq:proofE3}
d\!\left(\frac{\eta}{\xi^{1/2}}\right)
= -\frac{\eta}{2\xi^{3/2}}\, d\xi
= \frac{1}{2\eta}\left(-\frac{\eta^2}{\xi^{3/2}}\right) d\xi ,
\end{align}
where $\eta$ is treated as constant. Finally, multiplying (\ref{eq:proofE1}) by 
$d\xi$, taking into account (\ref{eq:proofE2}) and (\ref{eq:proofE3}), and integrating 
with the properly chosen limits, we obtain the desired decomposition, recovering 
Eq.\ (\ref{eq:altFlamm}).

Notice a somewhat peculiar way in which wormhole dynamics is described here. 
Due to our specific choice of spacetime slicing, the corresponding $\eta$-dependent 
embedding goes from being purely tachyonic (since the $\eta$-field is spacelike 
in the interval $-\infty < \eta <-1$, representing initially disconnected universes), 
to a mixure of timelike (for $0<r/r_{\rm s}< (1-\eta^2)/\eta^2$) and spacelike 
(for $r/r_{\rm s} > (1-\eta^2)/\eta^2$) when $\eta$ is restricted to the range 
$-1<\eta<1$ (the throat), and then back to purely tachyonic for $1<\eta<+\infty$ 
(again corresponding to disconnected universes). In principle, this description is 
sufficient for the problem at hand, since the $\eta$ and $\chi$ fields are never 
tangent to one another on the extended manifold. 

\section{Simplified approach}
\label{sec:simplifiedApproach}

Let us now explore the simplified approach mentioned in the Introduction, 
which assumes familiarity with the standard Kruskal--Szekeres line element, 
Eq.\ (\ref{eq:KSMetricStandard}). The key idea is to remove the exponential 
factor in the metric by introducing a compensating exponential in the coordinate 
transformation, Eq.\ (\ref{eq:TandX}), subject to
\begin{equation}
\label{eq:rSimplifiedDefinition}
\frac{r}{r_{\rm s}} = 1+\chi^2-\eta^2,
\end{equation}
which guarantees that Eq.\ (\ref{eq:transcendentalKS})---and, therefore, 
Einstein's field equations---are satisfied. This strategy is loosely analogous 
to the way a coordinate singularity in the metric can be canceled by a 
corresponding singularity in the transformation equations. The calculation 
then proceeds as follows:

From Eqs.\ (\ref{eq:TandX}) and (\ref{eq:rSimplifiedDefinition}) we get,
\BEq
dT=\left(\frac{dr}{2r_{\rm s}}\eta + d\eta\right)e^{r/(2r_{\rm s})},
\quad
dX=\left(\frac{dr}{2r_{\rm s}}\chi + d\chi\right)e^{r/(2r_{\rm s})},
\quad
\frac{dr}{r_{\rm s}}=2(\chi d\chi -\eta d\eta),
\EEq
which after squaring gives,
\BEq
dT^2=\left(\frac{dr^2}{4r_{\rm s}^2}\eta^2 
+ \frac{\eta}{r_{\rm s}}d\eta dr 
+ d\eta^2\right)e^{r/r_{\rm s}},
\quad
dX^2=\left(\frac{dr^2}{4r_{\rm s}^2}\chi^2 
+ \frac{\chi}{r_{\rm s}}d\chi dr 
+ d\chi^2\right)e^{r/r_{\rm s}},
\quad
\frac{dr^2}{4r_{\rm s}^2}=(\chi d\chi -\eta d\eta)^2.
\EEq
Substituting into Eq.\ (\ref{eq:KSMetricStandard}) produces (with the spherical 
part omitted for clarity),
\begin{align}
ds^2 
&=
\frac{4r_{\rm s}^3}{r} e^{-r/r_{\rm s}}
\left[
(\eta^2-\chi^2)\frac{dr^2}{4r_{\rm s}^2}
+ (\eta d\eta - \chi d\chi)\frac{dr}{r_{\rm s}}
+ d\eta^2 - d\chi^2
\right] e^{+r/r_{\rm s}}
\nonumber \\
&=
\frac{4r_{\rm s}^3}{r} 
\left[
(\eta^2-\chi^2)\frac{dr^2}{4r_{\rm s}^2}
+ (\eta d\eta - \chi d\chi)\frac{dr}{r_{\rm s}}
+ d\eta^2 - d\chi^2
\right] 
\nonumber \\
&=
\frac{4r_{\rm s}^3}{r} 
\left[
\left(1-\frac{r}{r_{\rm s}}\right) (\eta d\eta - \chi d\chi)^2
-2(\eta d\eta - \chi d\chi)^2
+ d\eta^2 - d\chi^2
\right] 
\nonumber \\
&=
\frac{4r_{\rm s}^3}{r} 
\left[
-\left(1+\frac{r}{r_{\rm s}}\right)(\eta d\eta - \chi d\chi)^2
+ d\eta^2 - d\chi^2
\right],
\end{align}
which is the same as our main Eq.\ (\ref{eq:v04compact}). Generalization may 
then be implemented via the transformation
\begin{equation}
T = e^{r/(2r_{\rm s})} \eta(x^0, x^1), 
\quad
X = e^{r/(2r_{\rm s})} \chi(x^0, x^1),
\end{equation}
subject to
\begin{equation}
\label{eq:rGeneral}
r(x^0, x^1) = r_{\rm s}\left(1+\chi^2-\eta^2\right),
\end{equation}
and
\begin{align}
\label{eq:metricGeneral}
ds^2 
&= \frac{4r_{\rm s}^3}{r} 
\left[
\left(\frac{dr}{2r_{\rm s}}\eta + d\eta\right)^2
-
\left(\frac{dr}{2r_{\rm s}}\chi + d\chi\right)^2
\right],
\end{align}
where $\eta$ and $\chi$ are now regarded as arbitrary functions of the new 
coordinates $x^0$ and $x^1$. Eq.\ (\ref{eq:rGeneral}) again guarantees the 
validity of the Einstein field equations. In a sense, this simple yet unconventional 
procedure brings the playful challenge of creating maximal extensions directly 
into the classroom. With a suitable choice of the areal radius $r(x^0, x^1)$, a bit 
of patience, and a bit of luck, anyone can now construct their own beautiful 
maximal extension.

\section{Conclusions}

We have presented a simple, physically motivated construction that provides 
further clarification of Fronsdal’s original, historically important embedding of 
the maximally extended Schwarzschild manifold in a higher-dimensional 
pseudo-Euclidean space. Along the way, we proposed a modified, four-dimensional 
form of Fronsdal’s metric, which may be viewed both as Israel’s line element recast 
into a highly symmetrical form and, in a certain sense, as a dual and complementary 
reformulation of the standard Kruskal-Szekeres system. The main attractive features 
of the proposed extension are: (1) the explicit form of the transformation equations 
connecting it to standard Schwarzschild coordinates; (2) the explicit form of the areal 
radius and the resulting four-dimensional metric, which may prove useful for general 
relativistic calculations; and (3) the regular appearance of the corresponding spacetime 
representation diagram.

Despite having been frequently cited---especially during the early years of black 
hole research---Fronsdal’s original work has not been widely adopted in educational 
literature on general relativity. This may be due to its formal and rather abstract nature, 
which tends to appeal more to mathematically oriented researchers in the field. In this 
pedagogical account, we have reformulated it as a concrete and more accessible 
alternative within a fully four-dimensional framework, which may be better suited 
for educational purposes, especially in situations where the emphasis is on explicit 
computations rather than on diagrammatic causal analysis.

\appendix

\section{Christoffel symbols and curvature}
\label{appendix:1}

Here, for completeness, we list non-vanishing Christoffel symbols
and curvature components associated with (\ref{eq:v04}):
\begin{align}
&
\Gamma^{0}_{00}
=
-\Gamma^{1}_{10}
=
\frac{\eta \left(\chi^2\left(1+\chi^2\right)-\left(1-\eta^2\right)^2\right)}
{\left(r/r_{\rm s}\right)^3},
\quad
\Gamma^{0}_{10}
=
-\Gamma^{1}_{11}
=
\frac{-\chi \left(\eta^2\left(1-\eta^2\right)+\left(1+\chi^2\right)^2\right)}
{\left(r/r_{\rm s}\right)^3},
\\
&
\Gamma^{0}_{11}
=
\frac{\eta \left(\eta^4-4 \eta^2 \chi^2-3 \eta^2+3 \chi^4+6 \chi^2+3\right)}
{\left(r/r_{\rm s}\right)^3},
\quad
\Gamma^{1}_{00}
=
\frac{-\chi \left(3 - 6 \eta^2 + 3 \eta^4 + 3 \chi^2 - 4 \eta^2 \chi^2 + \chi^4\right)}
{\left(r/r_{\rm s}\right)^3},
\\
&
\Gamma^{0}_{22}
=
-\frac{\eta}{2},
\quad
\Gamma^{1}_{22}
=
-\frac{\chi}{2},
\quad
\Gamma^{2}_{12}
=
\Gamma^{3}_{13}
=
\frac{2 \chi}{r/r_{\rm s}},
\quad
\Gamma^{2}_{20}
=
\Gamma^{3}_{30}
=
\frac{-2 \eta}{r/r_{\rm s}},
\\
&
\Gamma^{0}_{33}
=
-\frac{1}{2} \eta \sin ^2\theta ,
\quad
\Gamma^{1}_{33}
=
-\frac{1}{2} \chi \sin ^2\theta ,
\quad
\Gamma^{2}_{33}
=
-\sin\theta \cos \theta ,
\quad
\Gamma^{3}_{23}
=
\cot \theta ,
\end{align}
and
\begin{align}
&
R^{0}_{001}
= R^{1}_{110}
= \frac{4 \eta \chi \left(\eta^2-\chi^2-2\right)}{\left(r/r_{\rm s}\right)^4},
\\
&
R^{2}_{102}
=
R^{3}_{103}
=
R^{2}_{012}
=
R^{3}_{013}
=
\frac{2 \eta \chi \left(\eta^2-\chi^2-2\right)}{\left(r/r_{\rm s}\right)^4},
\\
&
R^{0}_{101}
=
\frac{4 \left(\chi^4-\left(\eta^2-2\right) \chi^2+1\right)}{\left(r/r_{\rm s}\right)^4},
\quad
R^{1}_{001}
=
\frac{4 \left(\eta^4-\eta^2 \left(\chi^2+2\right)+1\right)}{\left(r/r_{\rm s}\right)^4},
\\
&
R^{2}_{002}
=
R^{3}_{003}
=
\frac{-2 \left(\eta^4- \eta^2 \left(\chi^2+2\right)+1\right)}{\left(r/r_{\rm s}\right)^4},
\quad
R^{2}_{121}
=
R^{3}_{131}
=
\frac{-2 \left(\chi^4-\left(\eta^2-2\right) \chi^2+1\right)}{\left(r/r_{\rm s}\right)^4}
\\
&
R^{0}_{202}
=
R^{1}_{212}
=
\frac{-1}{2 (r/r_{\rm s})},
\quad
R^{3}_{232}
=
\frac{1}{r/r_{\rm s}},
\\
&
R^{0}_{303}
=
R^{1}_{313}
=
\frac{-\sin ^2\theta }{2 (r/r_{\rm s})},
\quad
R^{2}_{332}
=
\frac{-\sin ^2\theta }{r/r_{\rm s}},
\end{align}
where $x^0=\eta$, $x^1=\chi$, $x^2=\theta$, $x^3=\varphi$, 
and $r/r_{\rm s}=1+\chi^2-\eta^2$.

\section{The Central Observers}
\label{sec:centralObserver}

In this section we introduce freely falling ``central'' observers, 
the ones that populate the (time-dependent, spherical) boundary 
separating the two {\it connected} universes. On each of the 
embedding diagrams shown in Fig.\ \ref{fig:5}, central observers 
are represented by a point with $Z=0$ (the narrowest point of 
wormhole's neck), while in both standard Kruskal and modified 
diagrams (Fig.\ \ref{fig:1}) they are represented by the same 
vertical geodesic, $\chi=0$, $-1<\eta<1$, that runs from past to 
future singularity. We use the plural mode of description here 
because, formally, there is an infinite number of central observers 
differing from each other by their location on the boundary 
sphere. On the other hand, the Fermi normal coordinates 
introduced below should be thought of as pertaining to just 
one, specifically chosen, central observer. Our interest in central 
observers stems primarily from their simplicity, but also from 
their potential use for the description of physics in ``the central 
region''  (in 't Hooft's terminology \cite{thooft2024}) of the 
extended spacetime manifold.

Let us re-write our metric (\ref{eq:v04}) to quadratic
order in Fermi normal coordinates (FNC) \cite{manasse1963}, 
$x^\a_{\rm F} =(t_{\rm c},x^j) \equiv (t_{\rm c},x,y,z)$, 
surrounding a central geodesic, ${\cal C}$, defined by the 
condition $\chi=0$ (here, $t_{\rm c}$ denotes the proper time 
measured along the geodesic). The general form of the 
metric in FNC is
\begin{align}
g^{\rm F}_{00}
&=
\eta_{00}-R^{\rm F}_{0j0k}(t_{\rm c}) \, x^{j}x^{k}
+{\cal O}(x^3),
\\
g^{\rm F}_{0\ell}
&=
-\frac{2}{3}R^{\rm F}_{0j{\ell} k}(t_{\rm c}) \, x^{j}x^{k}
+{\cal O}(x^3),
\\
g^{\rm F}_{{\ell} m}
&=
\eta_{{\ell} m}-\frac{1}{3}R^{\rm F}_{{\ell}jmk}(t_{\rm c}) \, x^{j}x^{k}
+{\cal O}(x^3),
\end{align}
so the task is to find the corresponding curvature 
components evaluated on ${\cal C}$. Since in our case 
the metric on ${\cal C}$ is given by
\BEq
\label{eq:metricCentral}
\left. ds^2 \right|_{\cal C}
=
4r_{\rm s}^2(1-\eta^2) d\eta^2
- \frac{4r_{\rm s}^2}{1-\eta^2} d\chi^2
- r_{\rm s}^2(1-\eta^2)^2 d\Omega^2,
\EEq
we have,
\BEq
\label{eq:geodesicEqCentral}
dt_{\rm c}=2r_{\rm s}\sqrt{1-\eta^2} \, d\eta ,
\EEq 
which, upon integration, gives the relation between the
proper and coordinate time,
\BEq
\label{eq:tauCentral}
t_{\rm c} = {r_{\rm s}} \left(\eta \sqrt{1-\eta^2}+\arcsin \eta\right)
\approx
2 r_{\rm s} \eta\left(1-\frac{\eta^2}{6}-\frac{\eta^4}{40}\right),
\EEq
with observer's total lifetime being
\BEq
t_{\rm c}^{\rm (total)}=\pi r_{\rm s}.
\EEq
The choice of an orthonormal frame parallel transported 
along the geodesic is also straightforward, and, on the 
basis of Eq.\ (\ref{eq:metricCentral}), may be taken to be
\begin{align}
\hat{e}_{(0)}
&=\left.\frac{\partial}{\partial t_{\rm c}}\right|_{\cal C}
=\left(\frac{1}{2r_{\rm s}\sqrt{1-\eta^2}},0,0,0\right),
\\
\hat{e}_{(1)}
&=\left.\frac{\partial}{\partial x}\right|_{\cal C}
=\left(0,\frac{\sqrt{1-\eta^2}}{2r_{\rm s}},0,0\right),
\\
\hat{e}_{(2)}
&=\left.\frac{\partial}{\partial y}\right|_{\cal C}
=\left(0,0,\frac{1}{r_{\rm s}(1-\eta^2)},0\right),
\\
\hat{e}_{(3)}
&=\left.\frac{\partial}{\partial z}\right|_{\cal C}
=\left(0,0,0,\frac{1}{r_{\rm s}(1-\eta^2)\sin \Theta}\right).
\end{align}
Then, following the standard prescription for constructing FNC 
(see \cite{manasse1963} for details; also, \cite{poisson2009}) 
and using the results of Appendix \ref{appendix:1}, after some 
algebra we find,
\begin{align}
R^{\rm F}_{0101} 
= 
2A(\eta),
\quad
R^{\rm F}_{0202} 
= 
R^{\rm F}_{0303} 
=-A(\eta),
\quad
R^{\rm F}_{1212} 
= 
R^{\rm F}_{1313} 
=A(\eta),
\quad
R^{\rm F}_{2323} 
=- 2A(\eta),
\end{align}
with
\begin{align}
A(\eta)\equiv \frac{1}{2r_{\rm s}^2 \left(1-\eta^2\right)^3},
\end{align}
and, correspondingly,
\begin{align}
\label{eq:FNCcentral1}
g^{\rm F}_{00}
&=
1+ A\left(y^2+z^2-2x^2\right),
\quad
g^{\rm F}_{11}
=
-\left[
1
+ (A/3) \left(y^2+z^2\right)
\right],
\\
\label{eq:FNCcentral2}
g^{\rm F}_{22}
&=
-\left[
1
+ (A/3)\left(x^2-2z^2\right)
\right],
\quad
g^{\rm F}_{33}
=
-\left[
1
+(A/3)\left(x^2-2y^2\right)
\right],
\\
\label{eq:FNCcentral3}
g^{\rm F}_{12}
&=
g^{\rm F}_{21}
=
(A/3)xy,
\quad
g^{\rm F}_{13}
=
g^{\rm F}_{31}
=
(A/3)xz,
\quad
g^{\rm F}_{23}
=
g^{\rm F}_{32}
=
-(2A/3)yz.
\end{align}
The Fermi metric can be written in more compact form by 
introducing spherical coordinates, 
$(t_{\rm c},r_{\rm c},\theta, \varphi)$, with the $x$ direction 
playing the role of the polar axis,
\begin{align}
y=r_{\rm c} \sin \Theta \cos \Phi,
\quad
z=r_{\rm c} \sin \Theta \sin \Phi,
\quad
x=r_{\rm c} \cos \Theta,
\end{align}
resulting in
\begin{align}
\label{eq:FNCcentralSpherical}
ds^2_{\rm F}
=
\left[1- A r_{\rm c}^2\left(3\cos^2 \Theta - 1\right)\right]dt_{\rm c}^2
- dr_{\rm c}^2
-\left(1+A r_{\rm c}^2/3\right)r_{\rm c}^2d\Theta^2
-\left[1+ \left(A r_{\rm c}^2/3\right)\left(3\cos^2 \Theta - 2\right)\right]
r_{\rm c}^2\sin^2\Theta d\Phi^2.
\end{align}

\begin{acknowledgments}

The author thanks Sergei Kopeikin and Loris Magnani for stimulating discussions. 

\end{acknowledgments}

\end{document}